\documentstyle[wstwocl,epsfig]{article}
\begin{document}

\textheight=9.2in

\bibliographystyle{unsrt}    

\newenvironment{comment}[1]{}{}
\newcommand{\st}{\scriptstyle}
\newcommand{\sst}{\scriptscriptstyle}
\newcommand{\mco}{\multicolumn}
\newcommand{\epp}{\epsilon^{\prime}}
\newcommand{\ve}{\varepsilon}
\newcommand{\ra}{\rightarrow}
\newcommand{\ppg}{\pi^+\pi^-\gamma}
\newcommand{\vp}{{\bf p}}
\newcommand{\ko}{K^0}
\newcommand{\kb}{\bar{K^0}}
\newcommand{\al}{\alpha}
\newcommand{\ab}{\bar{\alpha}}
\def\be{\begin{equation}}
\def\ee{\end{equation}}
\def\bea{\begin{eqnarray}}
\def\eea{\end{eqnarray}}
\def\CPbar{\hbox{{\rm CP}\hskip-1.80em{/}}}

\def\ap#1#2#3   {{\em Ann. Phys. (NY)} {\bf#1} (#2) #3.}
\def\apj#1#2#3  {{\em Astrophys. J.} {\bf#1} (#2) #3.}
\def\apjl#1#2#3 {{\em Astrophys. J. Lett.} {\bf#1} (#2) #3.}
\def\app#1#2#3  {{\em Acta. Phys. Pol.} {\bf#1} (#2) #3.}
\def\ar#1#2#3   {{\em Ann. Rev. Nucl. Part. Sci.} {\bf#1} (#2) #3.}
\def\cpc#1#2#3  {{\em Computer Phys. Comm.} {\bf#1} (#2) #3.}
\def\err#1#2#3  {{\it Erratum} {\bf#1} (#2) #3.}
\def\ib#1#2#3   {{\it ibid.} {\bf#1} (#2) #3.}
\def\jmp#1#2#3  {{\em J. Math. Phys.} {\bf#1} (#2) #3.}
\def\ijmp#1#2#3 {{\em Int. J. Mod. Phys.} {\bf#1} (#2) #3.}
\def\jetp#1#2#3 {{\em JETP Lett.} {\bf#1} (#2) #3.}
\def\jpg#1#2#3  {{\em J. Phys. G.} {\bf#1} (#2) #3.}
\def\mpl#1#2#3  {{\em Mod. Phys. Lett.} {\bf#1} (#2) #3.}
\def\nat#1#2#3  {{\em Nature (London)} {\bf#1} (#2) #3.}
\def\nc#1#2#3   {{\em Nuovo Cim.} {\bf#1} (#2) #3.}
\def\nim#1#2#3  {{\em Nucl. Instr. Meth.} {\bf#1} (#2) #3.}
\def\np#1#2#3   {{\em Nucl. Phys.} {\bf#1} (#2) #3.}
\def\pcps#1#2#3 {{\em Proc. Cam. Phil. Soc.} {\bf#1} (#2) #3.}
\def\pl#1#2#3   {{\em Phys. Lett.} {\bf#1} (#2) #3.}
\def\prep#1#2#3 {{\em Phys. Rep.} {\bf#1} (#2) #3.}
\def\prev#1#2#3 {{\em Phys. Rev.} {\bf#1} (#2) #3.}
\def\prl#1#2#3  {{\em Phys. Rev. Lett.} {\bf#1} (#2) #3.}
\def\prs#1#2#3  {{\em Proc. Roy. Soc.} {\bf#1} (#2) #3.}
\def\ptp#1#2#3  {{\em Prog. Th. Phys.} {\bf#1} (#2) #3.}
\def\ps#1#2#3   {{\em Physica Scripta} {\bf#1} (#2) #3.}
\def\rmp#1#2#3  {{\em Rev. Mod. Phys.} {\bf#1} (#2) #3.}
\def\rpp#1#2#3  {{\em Rep. Prog. Phys.} {\bf#1} (#2) #3.}
\def\sjnp#1#2#3 {{\em Sov. J. Nucl. Phys.} {\bf#1} (#2) #3.}
\def\spj#1#2#3  {{\em Sov. Phys. JEPT} {\bf#1} (#2) #3.}
\def\spu#1#2#3  {{\em Sov. Phys.-Usp.} {\bf#1} (#2) #3.}
\def\zp#1#2#3   {{\em Zeit. Phys.} {\bf#1} (#2) #3.}

\setcounter{secnumdepth}{2} 


\title{DOUBLE SCALING OF ANGULAR CORRELATIONS INSIDE JETS
\footnotemark}

\firstauthors{Wolfgang Ochs}

\firstaddress{Max-Planck-Institut f\"ur Physik,
        F\"ohringer Ring 6,
        D--80805 M\"unchen,
        Germany}

\secondauthors{Jacek Wosiek}

\secondaddress{Jagellonian University,
         Reymonta 4, PL 30-059 Cracow,
        Poland}

\twocolumn[\maketitle\abstracts{Angular correlations of partons have
been derived for high energy
jets
in QCD. Interesting new scaling properties with two redundant variables
(jet energy $P$ and jet opening angle $\Theta$) emerge which can be
tested within a parton-hadron duality picture. Recent results from LEP
support the scaling in $\Theta$; the scaling in both $P$ and $\Theta$
could be tested with jets from deep inelastic scattering or with
high $p_T$ jets.}]

\section{Introduction}

There are indications that there is a soft confinement
mechanism which does not lead to large rearrangements of
momenta during the hadronisation process, so that QCD predictions on
multiparton final states can be meaningfully compared with
predictions on hadronic final states\cite{BCM,QCD}. An extreme
possibility is the hypothesis of ``Local Parton Hadron Duality'' (LPHD)
\cite{LPHD}, according to which the parton cascade
\footnotetext{~~presented by W.O. at the International
Europhysics Conference
on High Energy Physics, Brussels, July 1995}

The energy spectra of particles gave
initial support for this idea, in that
they showed the same depletion for soft particles as did the spectra
of particles due to the soft gluon interference.\cite{LPHD,OPAL1}.
Also the effect of the running of $\alpha_s$ is clearly visible in the
energy spectra \cite{LO}. It is therefore important to test these ideas
for other quantities as well, in particular multiparticle observables.

Recently we have studied the properties of angular correlations
in jets within the double logarithmic approximation (DLA) of QCD
which provides the leading order results at high energies
\cite{OW}. We derived explicit expressions for various two- and
multiparton observables. Here we will report on some general
scaling and universality properties of our results which follow from
the QCD cascade with angular ordering.

Results partly overlapping with ours have been also obtained by
other groups using somewhat different theoretical approaches,
on angular correlations, in particular azimuthal correlations \cite{DMO}
and on multiplicity moments \cite{DD,BMP}.

\section{Theoretical scheme}

We start from an integral equation for the generating functional
in DLA \cite{FAD,QCD} from which the properties of the cascade
can be derived. It takes into account the leading collinear and
soft divergencies but neglects recoil effects. This yields
the correct high energy asymptotic predictions and we investigated
to what extent these predictions apply to present energies by comparing
with the more complete MC \cite{HERWIG} calculations.
We derive from the
generating functional the general $n$-parton cumulant correlation
function $\Gamma^{(n)}(\Omega_1,\ldots, \Omega_n)$ in the
spherical angles $\Omega_i$. Simpler quantities are obtained by
partial integration. Finally we checked the backward consistency
of our results by full integration, so our two parton correlation
$\Gamma^{(2)}$ integrates to the gobal multiplicity moment
$C_2=F_2-1=<n(n-1)>/<n>^2-1=1/3$, the well-known DLA result.

\section{Scaling results}

We consider the distribution $\rho^{(2)} (\vartheta_{12}, P,\Theta)$
in the relative angle $\vartheta_{12}$ between two partons inside
a cone of half opening angle $\Theta$ around the primary parton
of momentum $P$, also the quantity $\hat r =\rho^{(2)}/\bar n^2$
normalized by the average multiplicity $\bar n$ in the cone
or $r=\rho^{(2)}/\rho^{(2)}_{\rm norm}$, where $\rho^{(2)}_{\rm norm}$
is constructed from uncorrelated pairs (``event mixing'').
We find that this correlation function, after an appropriate
rescaling depends on the three variables $\vartheta_{12}$,
$P$ and $\Theta$
only through the single variable
$\varepsilon(\vartheta_{12},P,\Theta)$
\bea
\frac{\ln r(\vartheta_{12},\Theta,P)}{2\beta\sqrt{\ln\frac{P
\Theta}{\Lambda}}}&=&\omega
(\varepsilon, 2)-2\sqrt{1-\varepsilon}  \label{rscal}\\
\varepsilon &=&\frac{\ln \frac{\Theta}{\vartheta_{12}}}{\ln
                     \frac{P\Theta}{\Lambda}} \label{exp}
\eea
The function $\omega (\varepsilon, 2)$ determines the exponential growth
of $\rho^{(2)}$ and the second term in (\ref{rscal}) of $\rho_{\rm
norm}^{(2)}$.
The function $\omega(\ve, n)$ enters the $n$-parton correlation
function and is exactly calculable, also various approximate forms are
known \cite{OW}. The rescaling factor in (\ref{rscal}) determines the
exponential growth of multiplicity and can also be written as
$\ln\bar n=2\beta\sqrt{\ln\frac{P\Theta}{\Lambda}}$
where $\beta^2=4 N_c/(11 N_c/3-2N_f/3)$ and $\Lambda$ is
the QCD scale parameter. The variable $\varepsilon$ can
be viewed as a rescaled
angle  $\vartheta_{12}>\Lambda/P$
and therefore $\varepsilon < 1$.
The scaling law (\ref{rscal}) could then also be written as
\be
\frac{\ln r(\vartheta_{12},\Theta,P)}{\ln\bar n(\Theta,P)}=
\chi(\varepsilon).  \label{rnscal}
\ee
In a recent analysis by the DELPHI collaboration at LEP \cite{DELPHI}
this new scaling
property has been verified for jet opening angles in a range
$30^o <\Theta <60^o$, for smaller angles $\Theta\sim 15^o$
the scaling is violated; in this range the
sensitivity to the jet direction
(sphericity axis) becomes important (see Fig.~1).
\setlength{\unitlength}{0.7mm}
\begin{figure}
\begin{picture}(100,110)(0,1)
\centerline{\mbox{\epsfxsize5.4cm\epsffile{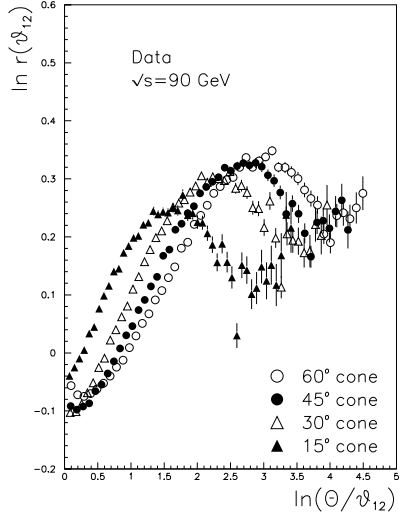}}}
\end{picture}
\vspace{-0.5cm}
\begin{picture}(100,110)(0,1)
\centerline{\mbox{\epsfxsize5.4cm\epsffile{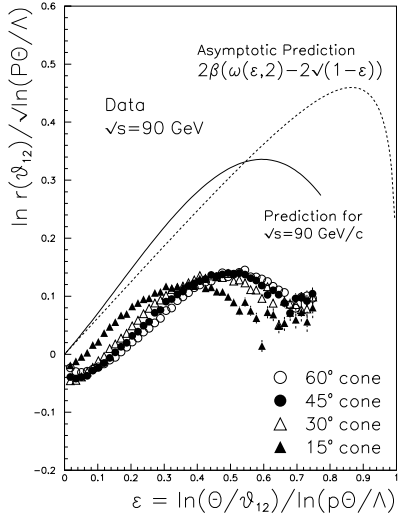}}}
\end{picture}
\vspace{-0.5cm}
\caption{(a) Correlation in the relative angle $\vartheta_{12}$ in the
forward cone of half opening angle
$\Theta$, for different $\Theta$ (b) the same, but in rescaled coordinates,
``$\varepsilon$ scaling'' (preliminary data from DELPHI).}
\end{figure}
The sensitivity to the jet direction
could be reduced by studying the corresponding
Energy-Multiplicity-Multiplicity Correlations (EMMC) in which
all particles serve in turn as jet axis with weight $E_i$ of
their energy \cite{EMMC}. Our analytical DLA predictions
reproduce roughly the shape of the distribution but not the
normalization which appears at nonleading order in the exponent.
The correlation $\hat r (\varepsilon)$ defined above has shown
better scaling properties in primary energy $P$ in our MC calculations.
The data \cite{DELPHI} show already a reasonable agreement with our
analytical predictions at present energies.

It will be very interesting to test the double scaling
in $\varepsilon$ by varying in the same
experiment the jet opening angle $\Theta$ and the jet energy $P$.
Recent results from HERA on energy spectra in the Breit frame
\cite{HERA} have been well described by QCD calculations as in the
$e^+e^-$-studies. The correlation data from deep inelastic
scattering and also from high $p_T$ jets are therefore well suited
to further establish the scaling properties of the underlying parton
cascade structure as derived from perturbative QCD.

Supported by PB 2P03B19609 and PB 2P30225206.


\section*{References}
\begin{thebibliography}{99}
\bibitem{BCM} A. Bassetto, M. Ciafaloni, G. Marchesini,
\prep{B100}{1983}{202}
\bibitem{QCD} Yu.L. Dokshitzer, V.A. Khoze, A.H. Mueller, S.I. Troyan,
{\em ``Basics of
    Perturbative QCD''}, Editions Fronti\`eres, Gif-sur-Yvette
    CEDEX-France (1991).
\bibitem{LPHD} Ya.I. Azimov, Yu.L. Dokshitzer, V.A. Khoze, S.I. Troyan,
\zp{C27}{1985}{65};~{\em Zeit. Phys} {\bf C31}\ (1961) 21.
\bibitem{OPAL1} M.Z. Akrawy et al. (OPAL coll.) \pl{247}{1990}{617}
\bibitem{LO} S. Lupia, W. Ochs, conf. paper eps0803, preprint
 hep-ph/9509249, to be publ.\ in {\em Phys. Lett} {\bf B}.
\bibitem{OW} W. Ochs, J. Wosiek, \pl{B289}{1992}{159}~
\pl{B304}{1993}{144} ~Preprint
hep-ph/9412384, to be publ.\ in {\em Zeit. Phys.} {\bf C}.
\bibitem{DMO} Yu. L. Dokshitzer, G. Marchesini, G. Oriani,
{\sl Nucl. Phys.} {\bf B387} (1992) 675.
\bibitem{DD} Yu. L. Dokshitzer, I. Dremin, \np{B402}{1993}{139}
\bibitem{BMP} Ph. Brax, J.L. Meunier, R. Peschanski,
\zp{C62}{1994}{649}
\bibitem{FAD} V.S. Fadin, {\em Yad. Fiz.} {\bf 37} (1983) 408.
\bibitem{HERWIG} G. Marchesini, B.R. Webber,
\np{B238}{1984}{1};~{\bf B310}~(1988)~461
\bibitem{DELPHI} F. Mandl, B. Buschbeck (DELPHI coll.), Proc. XXIV
  Int. Symp. on Multiparticle Dynamics, Vietri sul mare, 1994,
 Eds. A. Giovannini, S. Lupia, R. Ugoccioni, World Scientific,
Singapore (1995) p.52; conf. paper eps0553.
\bibitem{EMMC} Yu. Dokshitzer, V.A. Khoze, G. Marchesini, B.R. Webber,
\pl{B245}{1990}{243}
\bibitem{HERA} M. Derrick et al. (Zeus coll.) \zp{C67}{1995}{93}
S. Aid et al. (H1 coll) \np{445}{1995}{3}
\end{thebibliography}
\end{document}